\documentstyle[twocolumn,aps]{revtex}
\begin{document}

\input epsf
\twocolumn[\hsize\textwidth\columnwidth\hsize\csname
@twocolumnfalse\endcsname                         
\title{An effective lowest Landau level 
treatment of demagnetization in superconducting mesoscopic disks}
\author{J. J. Palacios}
\address{Departamento  de F\'{\i}sica Aplicada, 
Universidad de Alicante, San Vicente del Raspeig, Alicante 03080, Spain.}
\author{F. M. Peeters\cite{francois} and B. J. Baelus}
\address{Departement Natuurkunde, Universiteit Antwerpen (UIA), B-2610
Antwerpen, Belgium.}

\date{\today}
\maketitle

\widetext
\begin{abstract}
\leftskip 2cm
\rightskip 2cm

Demagnetization, which is inherently present in the magnetic response 
of small finite-size superconductors, 
can be accounted for by an effective $\kappa$  
within a two-dimensional lowest Landau level approximation of
the Ginzburg-Landau functional.
We show this by comparing the equilibrium 
magnetization of superconducting mesoscopic disks  
obtained from the numerical
solution of the three-dimensional Ginzburg-Landau equations 
with that obtained in the ``effective'' LLL approximation.
\end{abstract}

\vskip2pc]

\narrowtext

\section{Introduction}
The magnetic response of a three-dimensional superconductor depends on
intrinsic parameters such as
the coherence length $\xi$ and the penetration length $\lambda$
through the ratio $\kappa=\lambda/\xi$, but also depends
critically on the shape and orientation with respect to the applied
magnetic field $H$. The latter dependence makes it very difficult to extract
information  
about the true nature of the superconductor from magnetic measurements
and it can completely mask the intrinsic magnetic response. 
This phenomenon is known as demagnetization. The
best known example of how demagnetization affects the magnetic
properties of a finite-size superconductor is the appearance of the 
intermediate state\cite{Tinkham} in type-I superconductors. 
In very simple terms the intermediate state developes
due to the tendency for the expelled and ``overstretched'' magnetic flux lines
to penetrate back into the otherwise perfectly diamagnetic material.
For instance, macroscopic
type-I superconducting disks in perpendicular geometry or finite-length
cylinders with the rotational axis parallel to the field, which are 
still the focus of experimental and
theoretical studies\cite{Kuznetsov,Araujo,Brandt}, exhibit an 
apparent type-II magnetic response due to the formation of the
intermediate state.
Thin mesoscopic Aluminum disks (with radii ranging typically between 
0.1 and 2$\mu$m) have also received a lot of attention lately, in
part due to recent breakthroughs in magnetic\cite{Geim}
and resistive\cite{Moshchalkov} measurement techniques.  
Demagnetization in these mesoscopic disks is the focus of this paper.

From the theoretical point of view the demagnetization presented 
by macroscopic ellipsoidal samples 
can be simply accounted for by  a demagnetizing 
factor, $N$, which is a scalar if $H$ lies parallel to one
of the principal axis of the ellipsoid\cite{Tinkham}. $N$ runs from $N=1$ 
(maximum demagnetization) for the case of an infinite ellipsoid
perpendicular to $H$ to $N=0$ (zero demagnetization)
for the same ellipsoid parallel to $H$,
taking any value in between for intermediate situations like the
spherical geometry ($N=1/3$). When the samples are not ellipsoidal
the situation is more complicated. Still,
various effective models have been proposed to
account for demagnetizing effects in 
macroscopic samples with simple forms like strips, disks or 
finite cylinders with various orientations with
respect to the field\cite{Araujo,Brandt}, and
very accurate analytical results have been obtained in certain 
limits\cite{Brandt}. 
Compared to macroscopic disks, the situation for mesoscopic
disks\cite{Geim} gets complicated by the fact that
$\lambda$ can be comparable to the dimensions of the sample; in
particular, to the dimension parallel to the field or thickness of the
disk, $d$.  When this is the case the penetration
length is naively expected to get renormalized to
$\tilde\lambda=\lambda^2/d$\cite{Pearl} and the new $\kappa$
can be larger than $1/\sqrt{2}$, value that separates type-I from
type-II behavior.  Whether or not thin mesoscopic Aluminum disks exhibit 
truely type-II behavior and vortices form in the condensate can only
be unambiguously answered with imaging techniques\cite{Pannetier} since
demagnetization is always present and complicates the interpretation of the 
experimental results\cite{Geim}.

In principle, this non-trivial interplay between dimensions, geometry,
and intrisic parameters can only be addressed theoretically within a 
three-dimensional Ginzburg-Landau framework:
\begin{equation}
\frac{1}{2m^*}\left(-i\hbar{\vec\nabla} -
\frac{e^*}{c}{\vec{A}(\vec{r})}\right)^2\Psi(\vec{r}) =
-\alpha \Psi(\vec{r}) -\beta \Psi({\vec{r}}) |\Psi(\vec{r})|^2, 
\label{gl1}
\end{equation}   
\begin{equation}
\vec\nabla \times \vec\nabla \times \vec{A}(\vec{r}) = \frac{4\pi}{c}\vec{j},
\label{gl2}
\end{equation}   
where $\vec{j}$ is the superconducting current\cite{Tinkham}. 
These equations are 
obtained after extremizing the phenomenological 
Ginzburg-Landau energy functional:
\begin{eqnarray}
G_{\rm s}&=&G_{\rm n}+\int d\vec{r} \left[ \alpha |\Psi(\vec{r})|^2 +
\frac{\beta}{2}|\Psi(\vec{r})|^4 + \right.\nonumber \\
&&\left.\frac{1}{2m^*}\left|\left(-i\hbar\vec{\nabla} -
\frac{e^*}{c}{\vec A(\vec{r})}\right)\Psi(\vec{r}) \right|^2+
\frac{[h(\vec{r})-H]^2}{8\pi} \right],
\label{glf}
\end{eqnarray}   
where $G_{\rm n}$ is the Gibbs free energy of the normal state and $-i
\hbar\vec{\nabla} - e^* \vec{A}(\vec{r})/c$ is the momentum
operator for Cooper pairs of charge $e^*=2e$ and mass $m^*=2m$ in a
vector potential $\vec{A}(\vec{r})$ which is associated with a magnetic
induction $h(\vec{r})$. $\Psi(\vec{r})$ is the Cooper pair wave
function or complex order parameter and  the coefficients $\alpha$ and $\beta$ 
have the
usual values\cite{Tinkham} which scale 
phenomenologically with temperature\cite{Tinkham}.

The difficult task of solving the coupled Ginzburg-Landau differential
equations (\ref{gl1}) and (\ref{gl2}) for three-dimensional mesoscopic disks
has been undertaken numerically by 
Peeters and co-workers\cite{Peeters} (see also
Ref.\onlinecite{Zharkov} for the equivalent problem of cylinders). While,
typically, the order parameter within the disk
can be considered uniform in the direction of the field, i.e., 
effectively two-dimensional, the
three-dimensional magnetic flux structure needs to be taken fully into
account.
Considering the expected limitations of the effective Ginzburg-Landau
theory, the results\cite{Peeters} reproduce to a large extent
the experimental findings\cite{Geim}. Akkermans and co-workers have undertaken
similar studies from a different point of view. 
They have been able to minimize analytically the Ginzburg-Landau
functional close to the dual point $\kappa=1/\sqrt{2}$\cite{Akkermans1} and in 
the London limit $\kappa \rightarrow \infty$\cite{Akkermans2},
although demagnetizing
effects have been considered only at a phenomenological level.

The aim of this work is to show that for small disks there is 
an effective way to incorporate demagnetization in the solution of
the Ginzburg-Landau equations
without considering in detail the three dimensional structure of the magnetic 
field lines.  Essentially, we do this by 
introducing an effective $\kappa$, which we will denote by $\tilde\kappa$, 
that is geometry dependent.
Adequately chosen, the use of $\tilde\kappa$ instead of the 
intrinsic $\kappa$ can account, in part, for
demagnetizing effects. This is conveniently done within the
framework of the 
lowest Landau level (LLL) approximation which has been broadly used in the
case of bulk superconductors\cite{Tesanovic,MacDonald} and mesoscopic
systems\cite{Palacios,Peeters:prl}.
It will turn out that this procedure works rather well if we define 
two different
effective $\kappa$, one for the low 
magnetic field region, i.e., the Meissner state, and
one for magnetic fields near $H_{c2}$. 

In Sec. \ref{II} we
briefly review how the standard LLL approximation
must be modified to obtain the magnetic reponse of finite-size
superconductors where surface effects are dominant. We 
also compare our LLL approximation with traditional\cite{Abrikosov}
results for magnetization in the case of bulk systems.
In Sec. \ref{III} we present results for the equilibrium 
magnetization of mesoscopic disks obtained from the full numerical
solution of the Ginzburg-Landau differential equations\cite{Peeters}.     
We finally introduce the effective $\kappa$  in our LLL
approximation and make a comparison between the exact results and those
obtained in our ``effective'' LLL approach.

\section{Magnetization in the lowest Landau level approximation}
\label{II}
\subsection{The basics}
A fully two-dimensional alternative to the approaches mentioned in the
introduction for solving the Ginzburg-Landau equations
has been proposed by one of the authors\cite{Palacios}. 
One expands the order parameter in a set
of functions that are the lowest level solutions of the linearized version 
of Eq. (\ref{gl1}): 
\begin{equation}
\Psi(\vec{r})=\sum_{L=0}^{\infty} C_L \Upsilon_L(\vec{r})
 =\sum_{L=0}^{\infty} C_L \frac{1}{\sqrt{2\pi}}
e^{-iL\theta}\Phi_L(r).
\label{expansion}
\end{equation} 
It is essential for
the radial part of the expansion functions, $\Phi_L(r)$, to obey the
standard boundary conditions  of zero current through the
surface\cite{deGennes}. The presence of the boundary
implies  that, for large enough angular momentum $L$,
an infinite number of bulk Landau levels participate
(numerical work is usually required at this point). Nevertheless, 
we will still refer to this expansion as a LLL expansion. 
The other central idea behind this approximation is to consider the magnetic
induction to be spatially constant, $B$, but {\em not necessarily}
equal to the external field $H$, as it has been usually the case in
the context of the LLL approximation for bulk systems\cite{Tesanovic}. 
The magnetic induction $B$ sets the
scale of the magnetic length $\ell=\sqrt{e^*\hbar/cB}$ and, thus,
the scale of the expansion functions. Obviously, to consider a
constant value of $B$ across the superconductor is 
strictly valid only in the limit $B\rightarrow H$ and it does not
capture correctly the magnetic induction profile in the Meissner phase. 
On the other
hand, the LLL approximation is not expected to be valid below
$B\approx 0.3H_{c2}$, although this naive expectation 
has been shown to be higher than the real limit\cite{LLL}.
Still, as we will show in
the next section, this approach gives qualitative and quantitatively
good results for small and thin Aluminum disks in the whole range of fields as
long as $\tilde\kappa$ is adequately chosen. 

When one substitutes this expansion into the Ginzburg-Landau 
functional (\ref{glf}) one obtains an algebraic expression in terms of  the
complex coefficients $C_L$:
\begin{eqnarray}
G_{\rm s}&=&G_{\rm n}+\sum_{L=0}^{\infty} 
(\alpha+\epsilon_L) |C_L|^2 \nonumber \\
&&+ \frac{\beta}{2}
\sum_{L_1,L_2,L_3,L_4=0}^{\infty} C^*_{L_1}C^*_{L_2}C_{L_3}C_{L_4}
\int d{\vec r}\: \Upsilon^*_{L_1}\Upsilon^*_{L_2}\Upsilon_{L_3}\Upsilon_{L_4}
\nonumber
\\ && + V(B-H)^2/8\pi.
\label{LLL}
\end{eqnarray}     
The effect of the surface on the condensate is
cast in the Cooper pair energy $\epsilon_L$, which, unlike bulk
systems, presents a dip for states $L$ close to the
surface\cite{Palacios}, and in the ``interaction'' integrals.
In this form the Ginzburg-Landau functional can be easily minimized or, 
in general, extremized with respect to $C_{L_i}$ and $B$. 
This can even be done analytically when not
too many terms in the expansion (\ref{expansion})
participate\cite{Palacios,Peeters:prl}.

\subsection{Bulk magnetization}
It is illustrative at this point to consider a bulk system within our
LLL approximation. Let us assume an ellipsoidal form or a long
cylinder for which no demagnetization exists ($N=0$). 
In the spirit of Abrikosov's seminal work, it is
straightforward to obtain an expression for the
magnetic induction
\begin{equation}
B=H-\frac{H_{c2}-H}{2\beta_A\kappa^2 -1} ,
\end{equation}
and the free energy density
\begin{equation}
G_{\rm s}-G_{\rm n}/V=-\frac{(H_{c2}-H)^2}{8\pi(2\beta_A\kappa^2-1)}.
\end{equation}
In the above expressions $V$ is the volume
and we have made use of the Abrikosov parameter
$\beta_A= \langle \Psi(\vec{r})^2 \rangle^2/\langle \Psi(\vec{r})^4
\rangle$\cite{Abrikosov} which measures the uniformity of the
superconducting density.
Notice that the expression that we obtain for the magnetic induction 
is {\em different} from that obtained by
Abrikosov\cite{Abrikosov}:
\begin{equation}
\langle h(\vec{r})\rangle=H-\frac{H_{c2}-H}{2\beta_A\kappa^2 -\beta_A}.
\end{equation} 
The difference lies essentially in two facts. (i) 
We consider a uniform magnetic induction $B$ while, in Abrikosov's 
work, $h(\vec{r})$ is not uniform; instead, the spatial average
$\langle h(\vec{r})\rangle$ is calculated. (ii) Our LLL expansion
functions are evaluated for a value of $B$ that needs to be 
determined after minimization; on the other hand, Abrikosov's expansion 
functions are calculated at the upper critical field $H_{c2}$. 
These two significant 
differences account for the difference in the magnetization
expressions although, our approach is considerably simpler. Considering that 
when one minimizes with respect to the structure of the vortex lattice one 
obtains $\beta_A\approx 1.16$, which corresponds to a triangular 
arrangement of vortices\cite{Tinkham}, 
both approaches give a remarkably similar result for the bulk
magnetization.

\section{Demagnetization in the lowest Landau level approximation}
\label{III}
\begin{figure}
\centerline{\epsfxsize=8.0cm \epsfbox{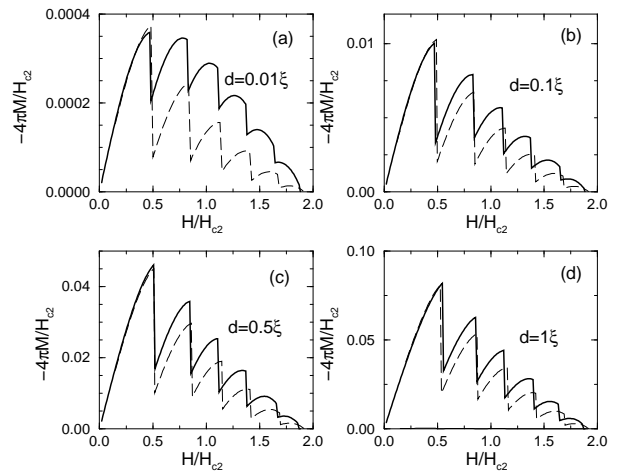}}
\caption{Equilibrium magnetization of a disk of radius $R=3\xi$ and 
bulk $\kappa=1$ for four different thicknesses:
$d/\xi = 0.01 (a), 0.1 (b), 0.5 (c), 1.0 (d)$.
Solid lines correspond to the full numerical
solution and dashed lines correspond to the LLL approximation using
different values of $\tilde\kappa$: (a) $\tilde\kappa=32$, (b) 
$\tilde\kappa=6.1$, (c) $\tilde\kappa=2.9$, and (d)
$\tilde\kappa=2.15$. }
\label{R=3}
\end{figure}

We now compare our LLL approximation  with the exact magnetization results
for mesoscopic disks of typical sizes $R\sim \xi$. Here,  only one
state $\Upsilon_L$ participates in the expansion (\ref{expansion})
for any value of the external field  or, in other words, 
the order parameter has
a well-defined value of the angular momentum and
forms a giant vortex\cite{Fink}. Minimizing Eq. \ref{LLL} with respect
to $C_L$ one obtains 
\begin{equation}
G_{\rm s}-G_{\rm n}=-\frac{[1-B\epsilon_L]^2}{\tilde\kappa^2 B R^2
\int \Phi^4}+(B-H)^2,
\label{1c}
\end{equation}  
where the minimal value of B
is obtained numerically.  The free energy is given
in units of $H_{\rm c2}^2V/8\pi$, $\epsilon_L(B)$ is given in units 
of the LLL
energy $\hbar\omega_c/2$ (with $\omega_c=e^*B/m^*c$), $R$ is
in units of $\xi$, and $B$ and $H$ are given
in units of $H_{\rm c2}$. Notice that in the denominator of the above 
expression we have written $\tilde\kappa$ instead of 
$\kappa$. As mentioned in the introduction,
finite thickness affects the real value of $\kappa$ through the
expression $\lambda^2/d$, but also does demagnetization in a more 
complicated way. We propose to consider $\tilde\kappa$ as an effective
parameter that depends on the geometry, i.e., on $R$ and $d$ in such a
way that the exact magnetic response is approximately
reproduced (within the limits
of the LLL approximation).  Figure \ref{R=3} shows 
the equilibrium\cite{Palacios:note} magnetization of a 
disk characterized by  $R=3\xi$ and $\kappa=1$
for four different values of the disk thickness: 
$d/\xi = 0.01 (a), 0.1 (b), 0.5 (c), 1.0 (d)$. 
Dashed lines correspond to the LLL
approximation and  solid lines to the three-dimensional 
numerical solution of Eqs. (\ref{gl1}) and (\ref{gl2}).  
Each jump
is associated with a transition from one 
giant vortex to another with one unit
more or less of vorticity ($L \rightarrow L \pm
1$)\cite{Palacios,Peeters,Peeters:prl}. The value of
$\tilde\kappa$ has been chosen to match the slope of the quasi-linear
magnetic response of the Meissner state ($L=0$).
As mentioned before, it is
in this state that our approximations are expected to be less 
accurate since the magnetic induction is clearly not uniform there;
still, keeping in mind that we have chosen the worst case
scenario for our fitting, it is a simple and unambiguous criterion for
choosing $\tilde\kappa$ and we will use it in what follows. In this figure
one can appreciate that the magnetization obtained in the LLL
approximation follows closely the ``exact'' magnetic response, both in
magnitude and position of the jumps. As the
thickness increases and the disk resembles more and more a long
cylinder, the value of $\tilde\kappa$ approaches the intrinsic one and
the effective LLL approximation works better.
Notice, however, how demagnetization manifests itself in that the
value of $\tilde\kappa$  [(a) $\tilde\kappa=32$, (b)
$\tilde\kappa=6.1$, (c) $\tilde\kappa=2.9$, and (d)
$\tilde\kappa=2.15$] is different from the one expected from
the simple scaling $\lambda^2/d$ (see Fig. \ref{kappa}).  This
manifestation is more clear for larger disks as will
be shown below.

\begin{figure}
\centerline{\epsfxsize=8.0cm \epsfbox{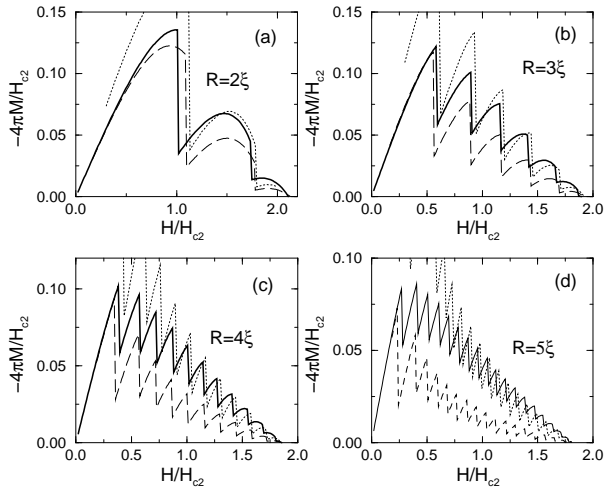}}
\caption{Equilibrium magnetization of a disk of thickness $d=0.1\xi$ 
and bulk $\kappa=0.28$ for different radii: 
$R\xi = 2.0 (a), 3.0 (b), 4.0 (c), 5.0 (d)$. 
Solid lines correspond to the full numerical
solution, dashed lines correspond to the LLL approximation using
different values of $\tilde\kappa$: (a) $\tilde\kappa=1.45$, (b) 
$\tilde\kappa=1.78$, (c) $\tilde\kappa=2.2$, and (d)
$\tilde\kappa=2.7$, and dotted lines correspond to the LLL approximation using
(a) $\tilde\kappa=1.20$, (b) 
$\tilde\kappa=1.35$, (c) $\tilde\kappa=1.45$, and (d)
$\tilde\kappa=1.55$}
\label{D=0.1}
\end{figure}            
Confident in the validity of the effective LLL approximation to
describe the actual magnetic response of small type-II
disks (we have repeated the analysis just described for 
different disk radii with similar  encouraging
results), we now try to push it further to include the description of
type-I mesoscopic disks like those of Geim's 
experiments\cite{Geim}.
Figure \ref{D=0.1} shows the equilibrium magnetic response of an
Aluminum disk ($\kappa=0.28$) of thickness $d=0.1\xi$ for four different
values of $R/\xi = 2.0 (a), 3.0 (b), 4.0 (c), 5.0 (d)$. 
The same convention as before for fixing $\tilde\kappa$
has been followed (dashed lines). 
Although for $d=0.1\xi$ a renormalized $\kappa$ is
already expected, it is clear from the results that, as the radius
increases, one needs to increase $\tilde\kappa$ [(a) $\tilde\kappa=1.45$, (b) 
$\tilde\kappa=1.78$, (c) $\tilde\kappa=2.2$, and (d)
$\tilde\kappa=2.7$] in order to match the
position of the magnetization jumps and, to some extent, the magnitude of 
them. The fact that we have to use higher values of $\tilde\kappa$
as the disk radius increases 
(whereas the thickness remains constant) constitutes again a manifestation of 
demagnetizating effects which decrease the magnetic response
of the disk as the radius grows bigger (i.e., the disk 
resembles more and more a thin film in perpendicular
geometry). Even for the larger
disk considered the agreement of the LLL results with the exact
solution is fairly satisfactory. However, now it becomes more 
difficult to match the overall magnitude
of the magnetization in the whole range of fields with a single value
of $\tilde\kappa$. At large fields, where the LLL approximation
is expected to work better, the agreement
can be improved considering a different set of values for  $\tilde\kappa$
(dotted lines). Unlike the previous case, the fitting criterion is 
a little ambiguous at high fields, 
but one can clearly appreciate the improvement. As 
Fig. \ref{kappa} shows, in 
all cases considered $\tilde\kappa$ behaves almost linearly with $R$.
One last caveat: 
Multiple-vortex states\cite{Palacios} could be stable in
the condensate of the larger disk, but we are not considering here 
such a possibility since it is not relevant for our discussion.

\begin{figure}
\centerline{\epsfxsize=8.0cm \epsfbox{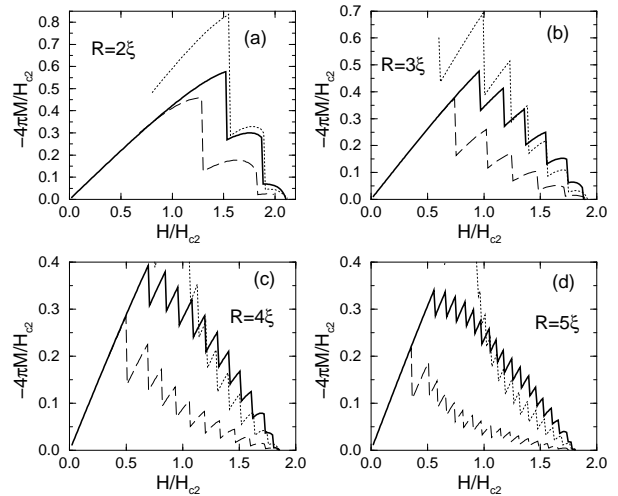}}
\caption{Equilibrium magnetization of a disk of thickness $d=0.5\xi$ 
and bulk $\kappa=0.28$ for 
different radii. Solid lines correspond to the full numerical
solution, dashed lines correspond to the LLL approximation using
different values of $\tilde\kappa$: (a) $\tilde\kappa=0.75$, (b) 
$\tilde\kappa=0.95$, (c) $\tilde\kappa=1.18$, and (d)
$\tilde\kappa=1.45$, and dotted lines correspond to the LLL approximation using
(a) $\tilde\kappa=0.55$, (b) 
$\tilde\kappa=0.65$, (c) $\tilde\kappa=0.70$, and (d)
$\tilde\kappa=0.75$}
\label{D=0.5}
\end{figure}

Finally, Fig. \ref{D=0.5} shows the equilibrium magnetization of an
Aluminum disk with $d=0.5\xi$ for different radii:
$R\xi = 2.0 (a), 3.0 (b), 4.0 (c), 5.0 (d)$. 
Now it becomes
more difficult to find an effective $\kappa$ that allows the LLL
approximation to reproduce, even if only qualitatively,
the exact results. This is not surprising
since, for such thickness, the type-I behavior of Aluminum manifests
itself more pronouncely. Still, the number of magnetization
jumps is approximately reproduced in all cases shown, but the  
the magnitude of the magnetization is appreciably
underestimated with the set of values for $\tilde\kappa$ that fit the
Meissner phase (dashed lines). Again, a different 
set can be chosen so that the
high-field response is approximately reproduced (dotted lines),
although the agreement is not completely satisfactory for large disks.

As conclusion, 
and for illustration purposes, we plot in Fig. \ref{kappa} 
$\tilde\kappa$ as a function of thickness and radius, including the cases
considered above. With the help of the calculations presented in this paper, 
we leave it to the reader to judge by himself when
the effective LLL can be safely used and  when and how
demagnetization effects can be accounted for in this  simple manner. 
It is beyond the scope of this paper 
to present a thourough study of all possible cases, but we expect this work to
serve as a useful guide to the specialized reader.

\begin{figure}
\centerline{\epsfxsize=8.0cm \epsfbox{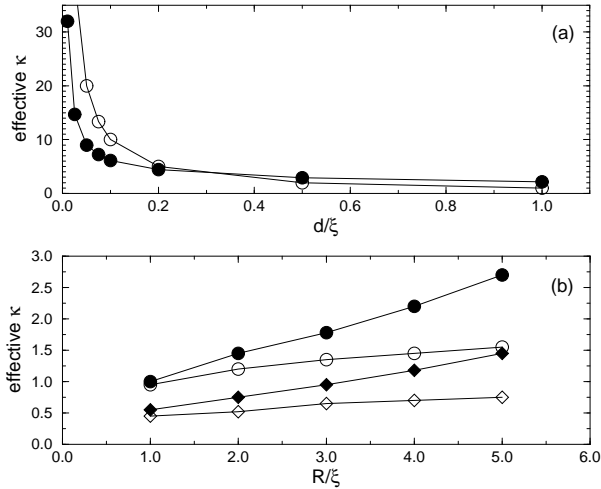}}
\caption{(a) Black dots: $\tilde\kappa$ as a function of the thickness for
$\kappa=1$ and $R=3\xi$. White dots: $\tilde\kappa$  expected from
the scaling $\lambda^2/d$. (b) $\tilde\kappa$ as a function of  $R$
for $\kappa=0.28$ and $d=0.1\xi$ (upper curves) and $d=0.5\xi$ (lower
curves). Black symbols denote fitting to the Meissner phase and white 
symbols to the
high-field magnetic response.}
\label{kappa}
\end{figure}            

\acknowledgments
Part of this work was supported by the Flemish Science Foundation (FWO-Vl),
the "Onderzoeksraad van de Universiteit Antwerpen", IUAP-IV, the European
ESF-project on Vortex Matter, the Spanish CICYT under Grant No. 1FD97-1358,
and the Generalitat Valenciana under Grant No. GV00-151-01.
Discussions with E. Akkermans, A. Geim, and K. Mallick
are gratefully acknowledged. JJP also acknowledges L. van Look for
discussions and hospitality.


\begin{references}
\bibitem{Palacios:note} Non-equilibrium magnetization curves 
(see Ref.\onlinecite{Palacios,Peeters}) could also be used for 
comparison purposes.
\end{references}
\end{document}